\documentclass[aps,prappl,reprint,longbibliography,superscriptaddress]{revtex4-1}  % for review and submission
\usepackage{graphicx}  % needed for figures
\usepackage{dcolumn}   % needed for some tables
\usepackage{multirow}
\usepackage{bm}        % for math
\usepackage{amssymb}   % for math
\usepackage{amsmath}   % for math
\usepackage{siunitx}   % for SI units
\usepackage{color}     % for \textcolor
\usepackage{float}     % for float placement

\usepackage[normalem]{ulem} % JB addition for striking out text

\usepackage{blindtext}

\usepackage[colorinlistoftodos]{todonotes}
\usepackage{verbatim}
\usepackage[normalem]{ulem} %Strikeout with \sout{} command

\DeclareSIUnit\torr{Torr}
\DeclareSIUnit\sq{\ensuremath{\Box}}
\DeclareSIUnit\dBm{dBm}

\begin{document}

\title{Stability of superconducting resonators: motional narrowing and the role of Landau-Zener driving of two-level defects}

\author{David~Niepce}
\affiliation{Chalmers University of Technology, Microtechnology and Nanoscience, SE-41296, Gothenburg, Sweden}
\author{Jonathan~J.~Burnett}
\affiliation{National Physical Laboratory, Hampton Road, Teddington, Middlesex, TW11 0LW, United Kingdom}
\author{Marina Kudra}
\affiliation{Chalmers University of Technology, Microtechnology and Nanoscience, SE-41296, Gothenburg, Sweden}
\author{Jared H. Cole}
\affiliation{Chemical and Quantum Physics, School of Science,
RMIT University, Melbourne VIC 3001, Australia}
\author{Jonas~Bylander}
\email{jonas.bylander@chalmers.se}
\affiliation{Chalmers University of Technology, Microtechnology and Nanoscience, SE-41296, Gothenburg, Sweden}
\vskip 0.25cm

\date{\today}

%% No more than 600 characters
\begin{abstract}
Frequency instability of superconducting resonators and qubits leads to dephasing and time-varying %quality factors 
energy-loss and hinders quantum-processor tune-up.
Its main source is dielectric noise originating in surface oxides.
Thorough noise studies are needed in order to develop a comprehensive understanding and mitigation strategy of these fluctuations.
Here we use a frequency-locked loop to track the resonant-frequency jitter of three different resonator types---one niobium-nitride superinductor, one aluminium coplanar waveguide, and one aluminium cavity---and we observe strikingly similar random-telegraph-signal fluctuations.
At low microwave drive power, the resonators exhibit multiple, unstable frequency positions, which for increasing power coalesce into one frequency due to motional narrowing caused by sympathetic driving of individual two-level-system defects by the resonator.
In all three devices we probe a dominant fluctuator,  %using Allan-deviation analysis. The fluctuator's 
finding that its amplitude saturates with increasing drive power, but its characteristic switching rate follows the power-law dependence of quasiclassical Landau-Zener transitions. 
\end{abstract}

%\pacs{}
%\keywords{}
\maketitle

\section{Introduction}
Superconducting microwave resonators~\cite{Zmuidzinas2012Feb}, in a variety of geometries, are essential tools in circuits for quantum computing~\cite{McRae2020}, microwave quantum optics~\cite{GU2017}, low-noise amplifiers~\cite{Aumentado2020}, radiation detectors~\cite{Zmuidzinas2012Feb}, and particle accelerators~\cite{Romanenko2017Dec,Romanenko2020Mar}. 
While the reduction of energy loss of resonators and qubits has received significant attention~\cite{Zmuidzinas2012Feb,McRae2020,Muller2019Oct}, leading to long-lived qubits~\cite{yan2016flux,Burnett2019Jun} and high-quality resonators~\cite{Bruno2015May},
%While significant attention has been directed to the study of energy loss~\cite{McRae2020}, 
far fewer studies report on parameter fluctuations~\cite{Muller2015Jul,Klimov2018Aug,Burnett2019Jun,Schlor2019Nov}.
Such fluctuations present a challenge to the bring-up and calibration stability of current quantum processors~\cite{Klimov2020}.
Thorough noise studies are needed in order to understand and mitigate these fluctuations.
%Thorough noise studies are needed in order to develop a comprehensive understanding of these fluctuations and to mitigate them.
%
Here, we examine the low-frequency jitter of three different types of superconducting resonator with the same experimental setup %by means of a Pound frequency-locked loop, 
and observe strikingly similar 
random telegraph signal (RTS) fluctuations.
%We identify individual random telegraph signal (RTS) fluctuations at several time scales in the noise, from 1~ms to 100~s.
At low excitation power, the RTS lead to multiple quasi-stable frequency positions that coalesce at high powers, 
%At low excitation power, the resonators exhibit multiple, unstable frequency positions, which for increasing power coalesce into one stable frequency.
which we interpret as motional narrowing caused by direct (sympathetic) driving of individual two-level system (TLS) defects by the resonator field, causing Landau--Zener transitions between the TLS states.
%

%%TC:ignore
\begin{figure*}[ht]
    \includegraphics[width=13.6cm]{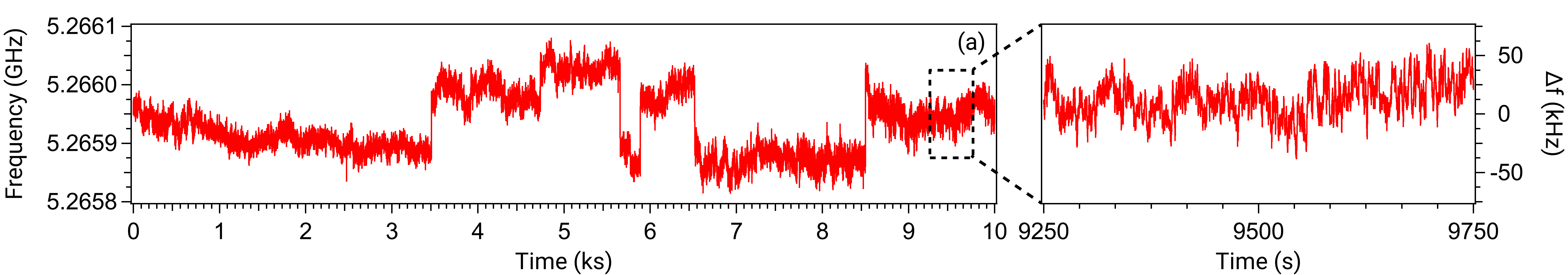} %Note fig size changes to 08*17cm
    \vskip 0.5em
    \includegraphics[width=4.48cm]{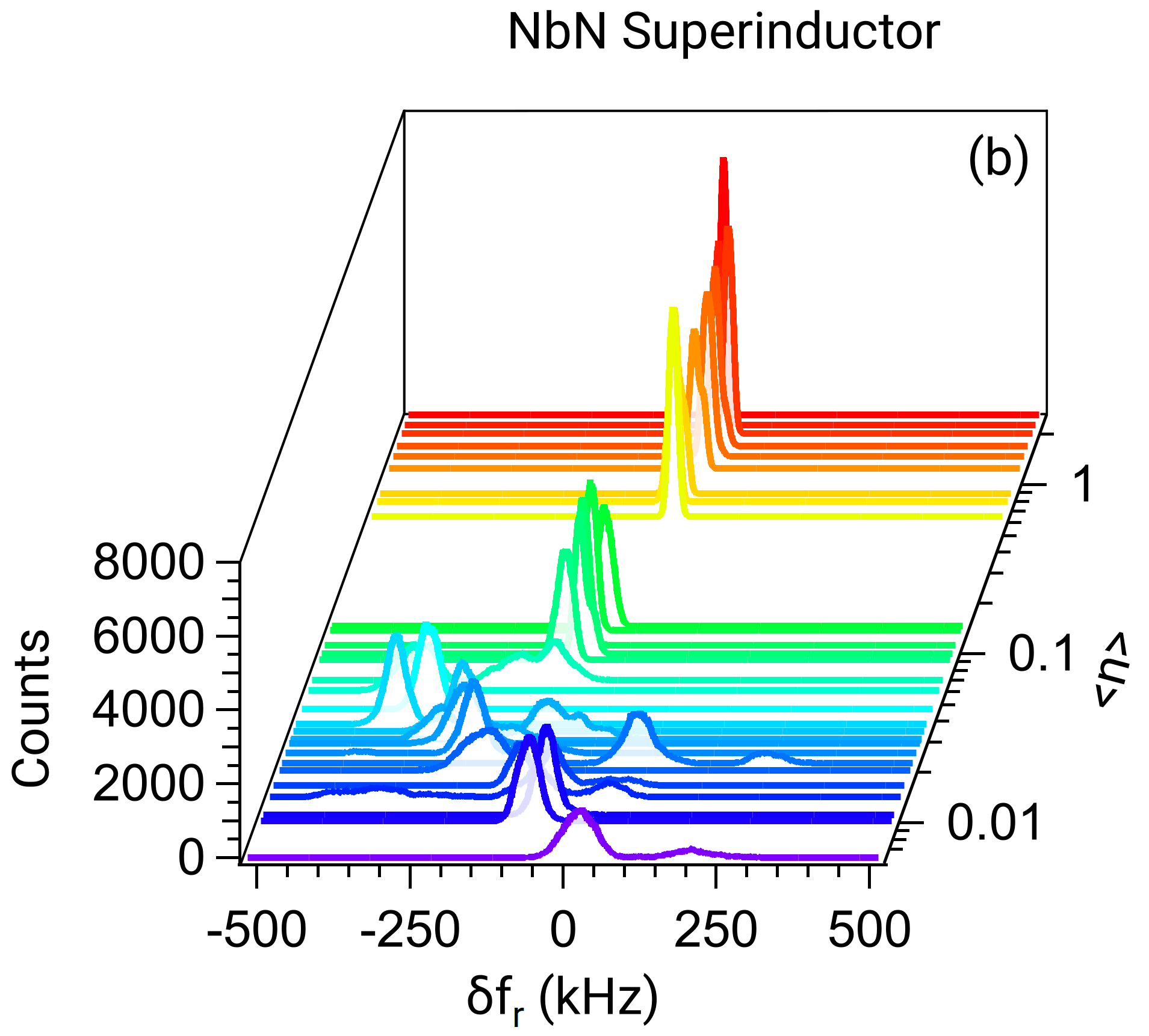}%Note fig size changes to 08*5.6cm
    \includegraphics[width=4.48cm]{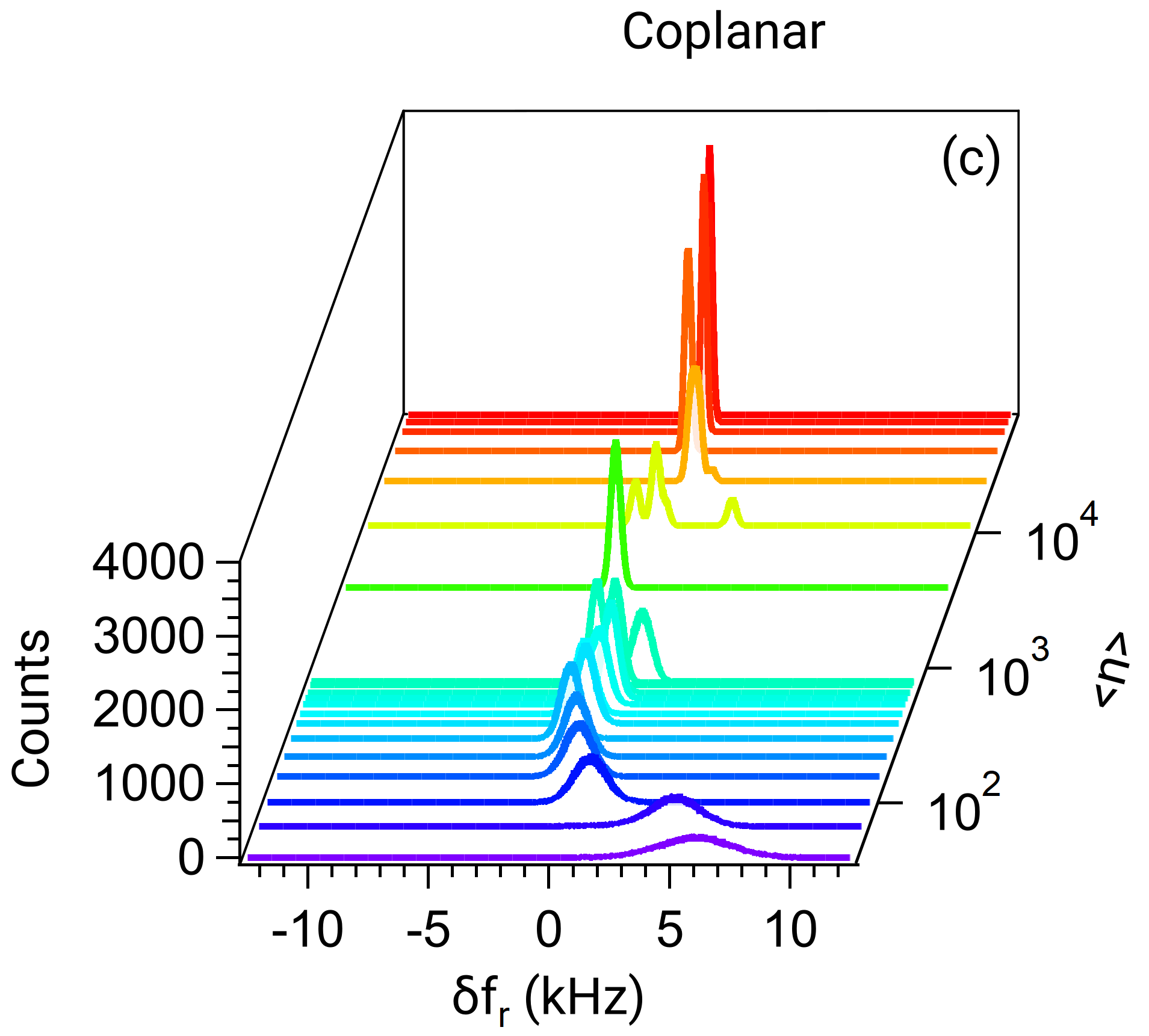}%Note fig size changes to 08*5.6cm
    \includegraphics[width=4.48cm]{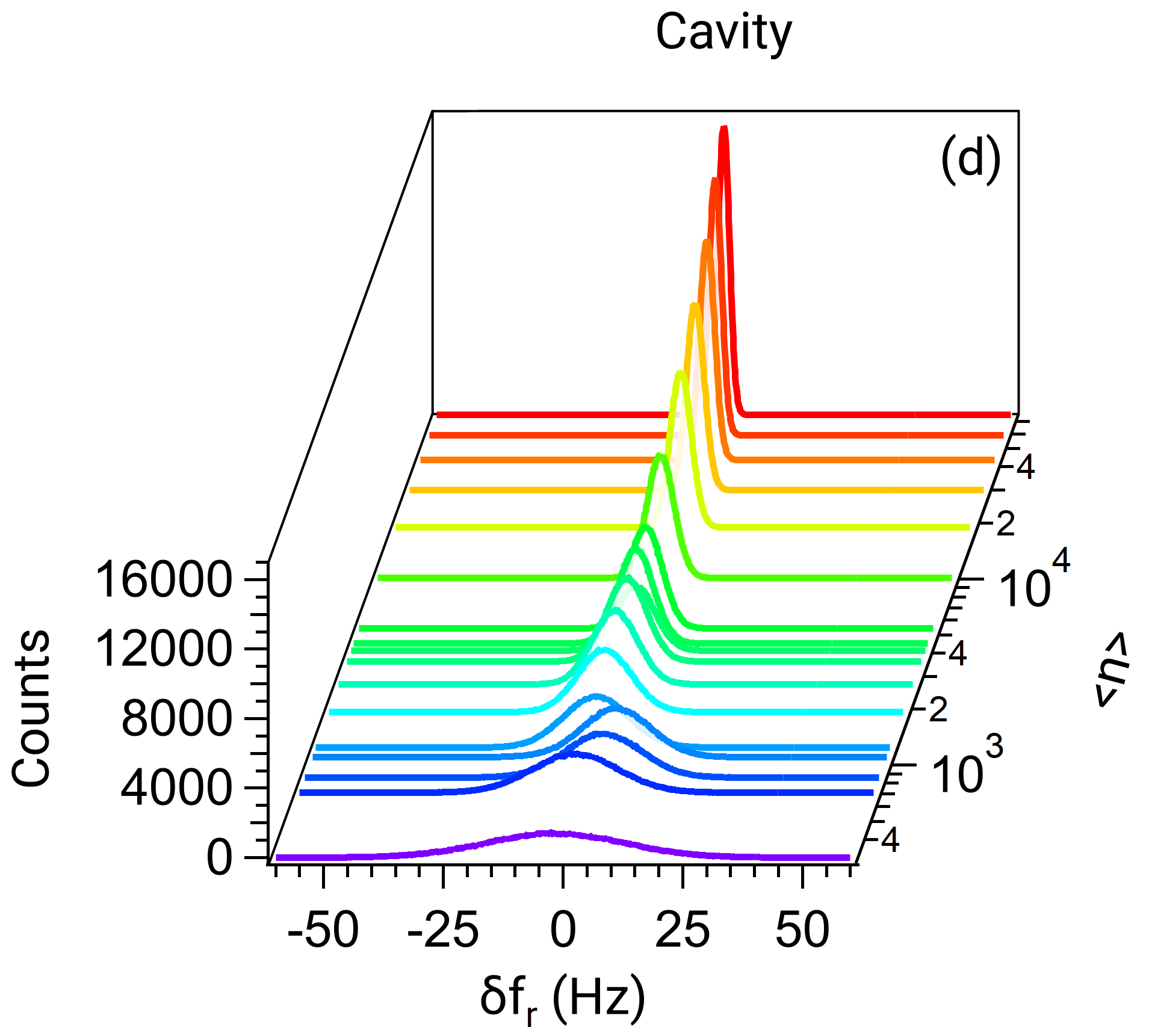}%Note fig size changes to 08*5.6cm
    
    \includegraphics[width=4.48cm]{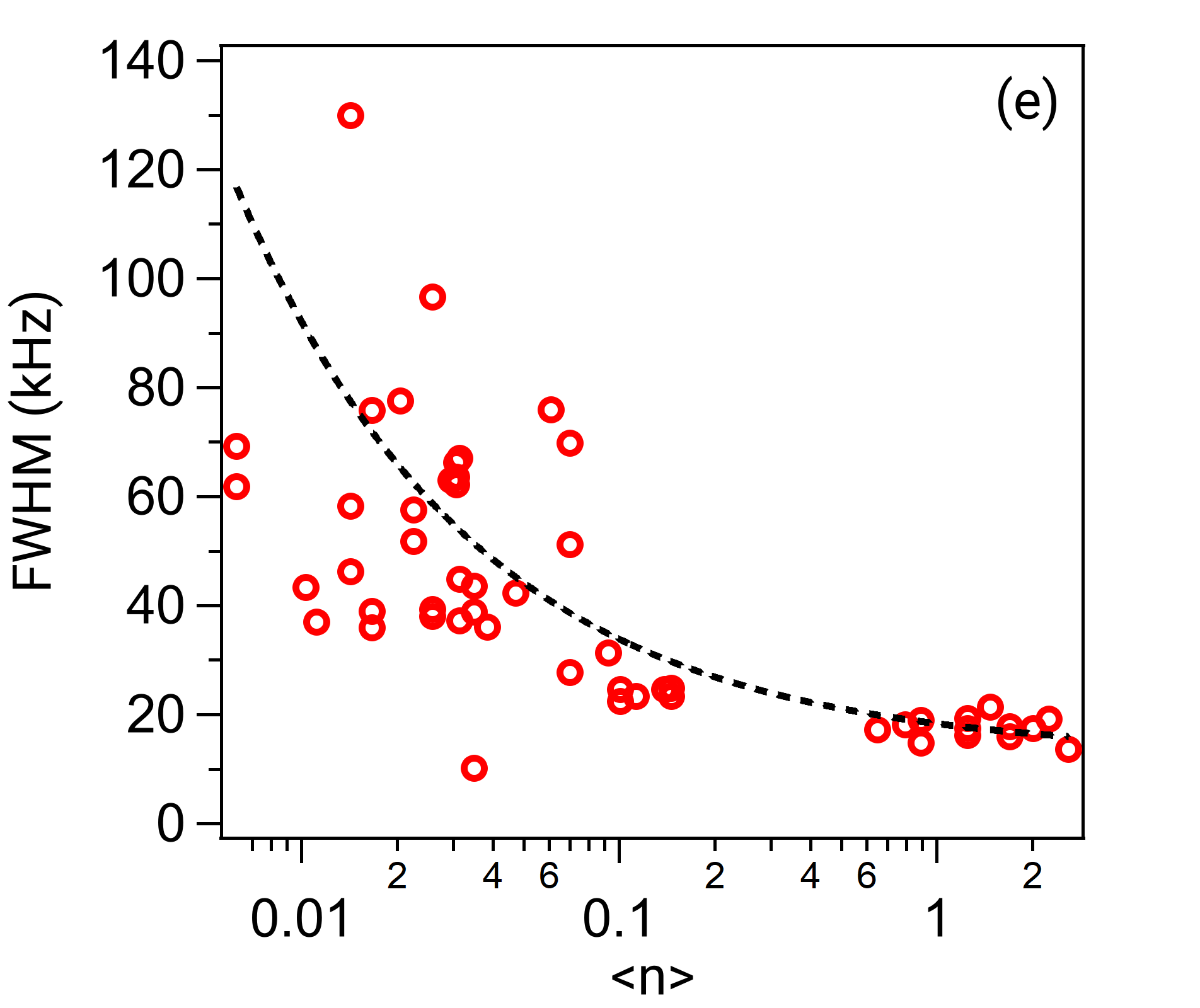}%Note fig size changes to 08*5.6cm
    \includegraphics[width=4.48cm]{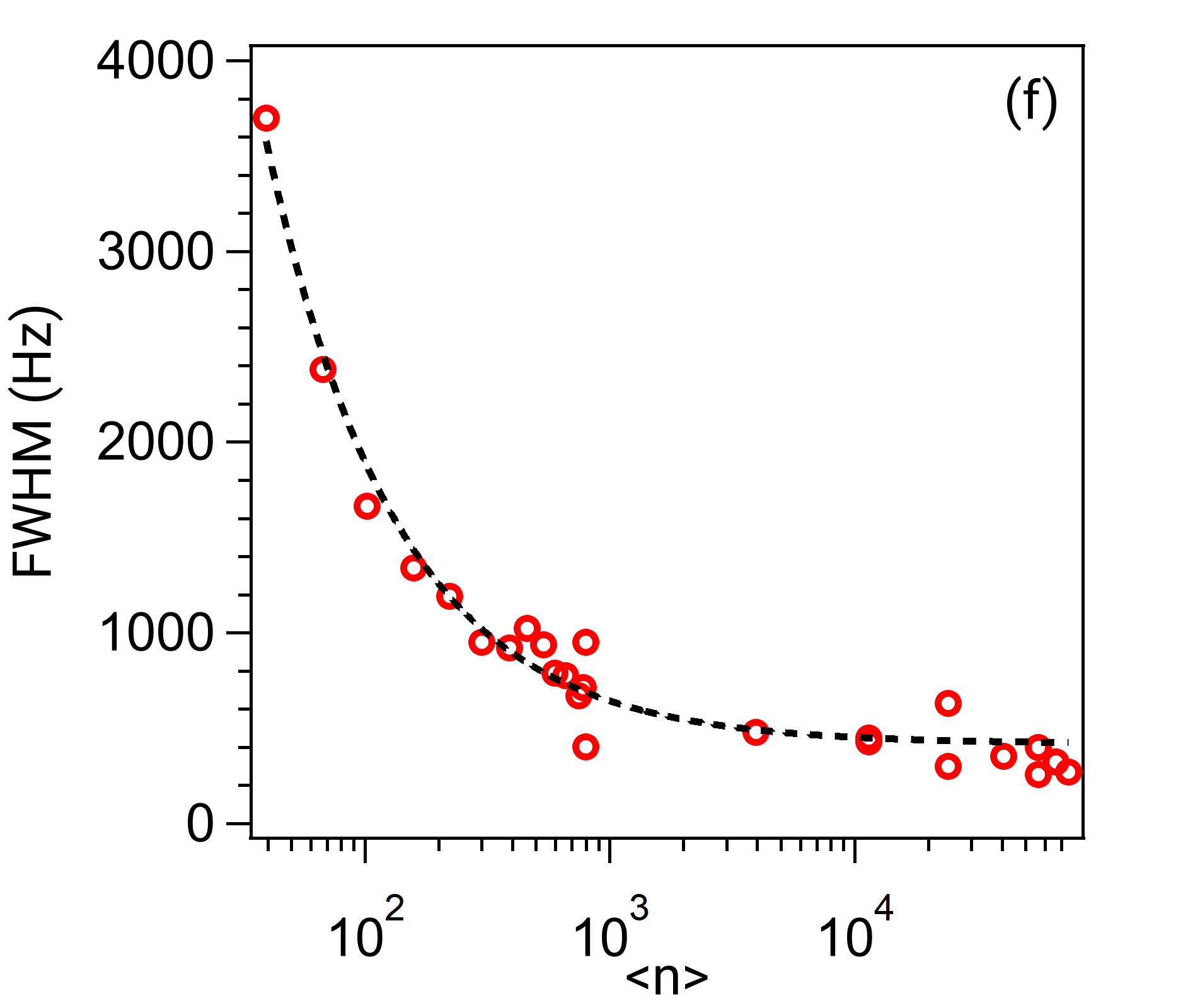}%Note fig size changes to 08*5.6cm
    \includegraphics[width=4.48cm]{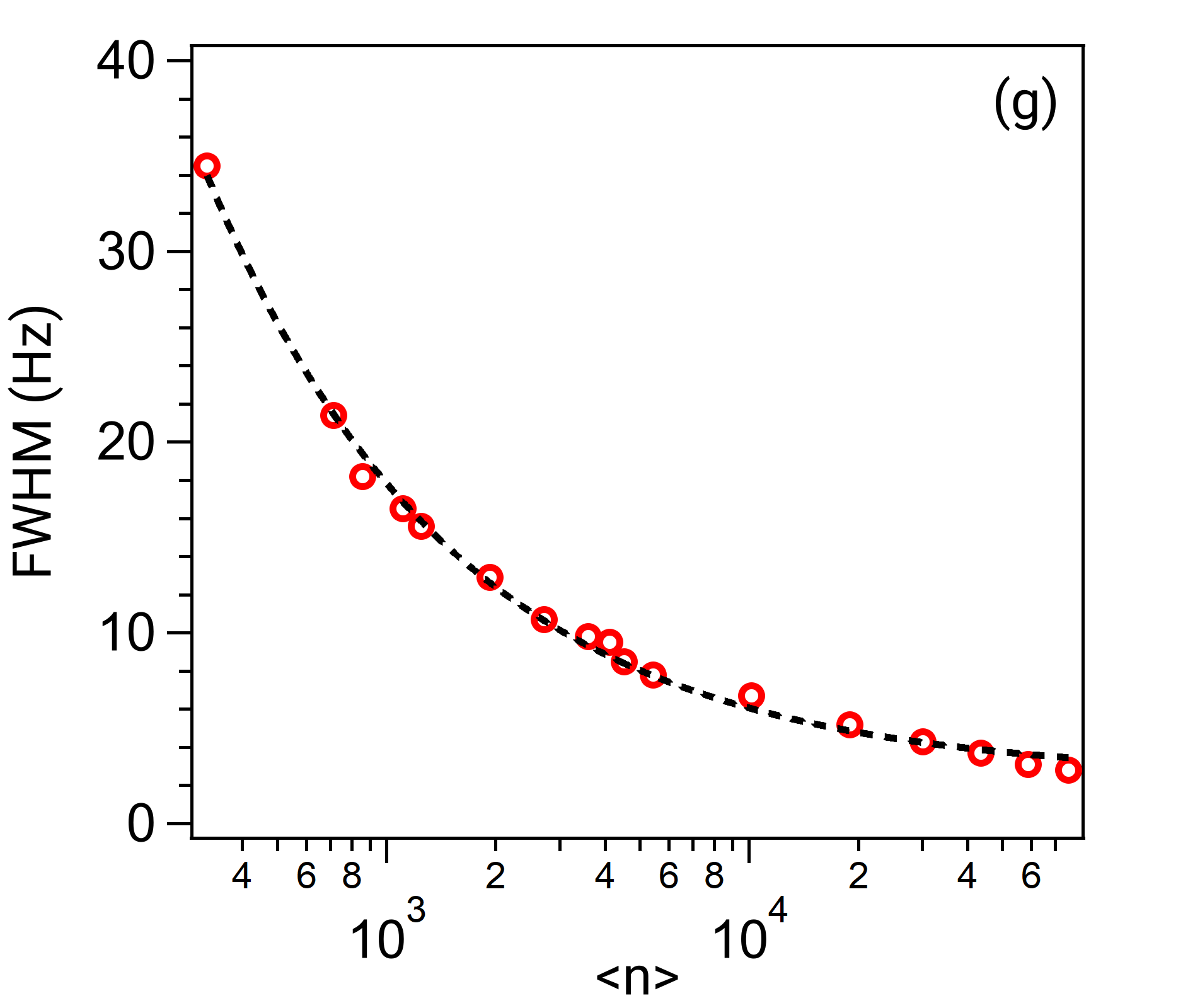}%Note fig size changes to 08*5.6cm

    \caption{\label{fig:waterfall} 
    \textbf{(a)} Raw frequency jitter of the nanowire resonator sampled at $\SI{100}{\hertz}$, at an applied power 
    %$P = \SI{-110.2}{\dBm}$ 
    corresponding to an average number of $\left< n \right> \simeq \SI{3e-2}{}$ photons in the resonator. 
    %$T = \SI{10}{\milli\kelvin}$.
    \textbf{(b--d)} Histograms of the frequency fluctuations for the three resonators vs.\@ applied power. The data is normalised to the mean frequency of the highest applied power. 
    \textbf{(e--g)} Peak widths (\small{FWHM}\normalsize) of the data in (b--d). (Note that \small{FWHM}\normalsize{} refers to the width of one peak in the histogram and not to the distance between resolvable peaks that correspond to quasi-stable configurations.)
}
\end{figure*}
%%TC:endignore

While the community agrees on the many underlying decoherence mechanisms which contribute to %limiting the coherence
decoherence, it remains divided on the relative importance of each mechanism. 
For example, the dissipation within Al resonators has been separately found to be limited by free-space-photon generated quasiparticles~\cite{Barends2011Sep} and two-level defects~\cite{Quintana2014Aug}. Similarly, dissipation in granular aluminium oxide resonators has been separately found to be limited by non-equilibrium quasiparticles~\cite{Grunhaupt2018Sep} and also by two-level defects~\cite{Zhang2019Jan}. Untangling these effects is complicated by experimental details that often differ: different signal filtering, use of infra-red absorber, magnetic shielding, and circuit-board enclosure vs.\ cavity enclosure. These differences make reports difficult to directly compare, resulting in conflicting interpretations of the underlying mechanism. This clearly demonstrates the need for experiments with common experimental details and for the standardization of measurement techniques.

Here we specifically use an identical %cryogenic, filter and measurement setup 
measurement and analysis infrastructure to %perform the same analysis on 
compare three very distinct types of superconducting resonators: an NbN ($T_{c} = 7.2$~K) 20~nm thick nanowire superinductor~\cite{Niepce2019Apr}, an Al ($T_{c} = 1.05$~K) 150~nm thick coplanar resonator, and finally an Al ($T_{c} = 1.18$~K) mm-scale 3D cavity resonator~\cite{Kudra2020Jun}. The device characteristics are summarised in Table~\ref{tab:devices} and in Methods. All three devices have similar resonant frequencies $f_r$ but vastly different superconducting properties, electric-field distributions, kinetic-inductance fractions, and internal quality factors $Q_i$. By performing %a
the same detailed analysis of the frequency `jitter' of these devices as a function of drive power we are able to directly compare the noise characteristics of all three devices.
%extract the noise characteristics of all three devices in a way that allows direct comparison between them.

A key observation is that the frequency response of these devices fluctuates as an RTS, i.e.\@ the frequency switches instantaneously between two or more discrete levels---see Fig.~\ref{fig:waterfall}(a). As the devices differ greatly in terms of design and dimensions, we attribute these fluctuations to two-level system (TLS) defects, %(two-level systems, TLS) which are intrinsic to the materials in the devices and which couple to the resonator and cause its frequency to fluctuate as a function of time. 
omnipresent in the dielectrics of superconductor surfaces and  interfaces. Dielectric \emph{loss}, due to near-resonant TLS, is a limiting factor for resonator internal quality factors, qubit relaxation times ($T_1$), MKID detection efficiencies~\cite{Zmuidzinas2012Feb}, and accelerator cavity efficacies~\cite{Romanenko2017Dec,Romanenko2020Mar}. Simultaneously, dielectric \emph{noise}, due to low-frequency TLS, leads to spectral instability, i.e.\@ fluctuations of $T_1$ (typically by 20\%) and of qubit frequencies (typically by a few kilohertz) with concomitant dephasing. 
The observed noise response reported here is entirely consistent with recent reports on fluctuations of single-TLS or few-TLS defects within superconducting qubits~\cite{Muller2015Jul,Klimov2018Aug,Burnett2019Jun,Schlor2019Nov}; however, in this setup, we are able to go further and identify the characteristics of a dominant TLS and even differentiate between device-specific response revealing TLS behaviour which is surprisingly consistent across devices. Analysis of the temporal fluctuations by spectral density and, particularly, by Allan-deviation techniques offers a window into the dynamics. As a result, we attribute the observed power dependence to sympathetic driving of the TLS bath by the resonator field.
Then, by analysing the fluctuations, we find that the RTS switching rate of all resonators follows a common power-law dependence that is consistent with the quasiclassical expression for the Landau-Zener transition rate.

\section{Results}

\subsection{Temporal frequency fluctuations}

%%TC:ignore
\begin{figure}[t]
    \includegraphics[width=8.5cm]{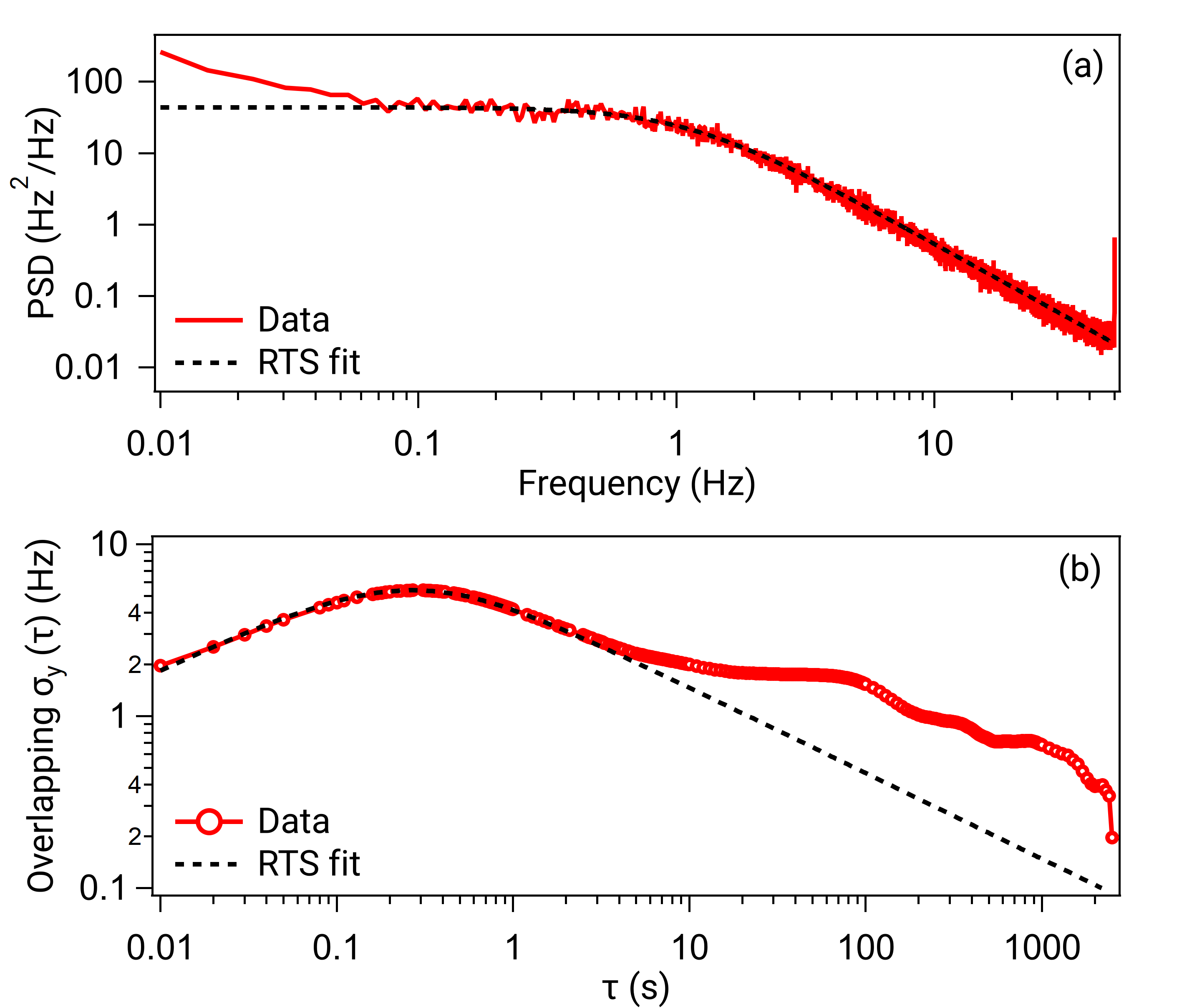}
    
    \caption{\label{fig:fits} Fitting of the noise to an RTS fluctuator model.
    The plots show an example of a Welch power spectral density $S_y(f)$ \textbf{(a)} and overlapping Allan deviation $\sigma_y(\tau)$ \textbf{(b)} for the measured frequency-fluctuation data from the cavity resonator held at $T = \SI{10}{\milli\kelvin}$ and with an applied microwave drive power $P = \SI{-131.5}{\dBm}$ ($\left<n\right> \simeq 715$). The data was sampled at $\SI{100}{\hertz}$. The dashed line corresponds to a fit of the  RTS fluctuator feature using a common set of fitting parameters for both traces  (eqs.~\ref{eq:lor_psd}--\ref{eq:lor_allan}). The data below 0.1~Hz (above 10~s) represents the tail of one or several secondary RTS fluctuators (see discussion in the main text).}
\end{figure}
%%TC:endignore

We use a Pound frequency-locked loop to measure the fluctuations of $f_r$ of the resonators for 2\,h\,45\,min (see Methods).  Figure~\ref{fig:waterfall}(a) shows an example of such a data set. We observe that the frequency fluctuates between discrete points, as is characteristic of an RTS. In fact, these fluctuations occur at all observable timescales as can be seen in the insert over a much shorter time period.

To qualitatively compare between the different devices, we calculate the histogram of frequency fluctuations measured on each of the resonators against circulating power in units of the average photon occupation number $\langle n\rangle$ (Fig.~\ref{fig:waterfall} (b--d)) and extract the histogram full width at half maximum ({\small{FWHM}}) (Fig.~\ref{fig:waterfall} (e--g)). We observe that the fluctuation amplitude (histogram width) is the highest for the nanowire resonator (b,e), lower in the coplanar resonator (c,f), and lowest in the cavity (d,g).
We attribute this to fluctuations of the real part of the dielectric susceptibility, which acts as an effective capacitance noise on the resonator and therefore leads to frequency fluctuations. Indeed, the nanowire has the highest sensitivity to electric fields, due to its very high impedance and %to the
high electric-field filling factor~\cite{Niepce2019Apr,Niepce2020Jan}. %of its surrounding dielectrics within the volume permeated by electric field, owing to its narrow geometry. 
In the coplanar resonator, the electric field is not as strongly coupled. Lastly, the cavity has the smallest filling factor and will therefore exhibit the least amount of frequency fluctuations.
We note that while the losses of superconducting cavities have been studied at sub-kelvin temperatures~\cite{Reagor2013May,Romanenko2017Dec,Romanenko2020Mar,Kudra2020Jun}, we have found no reports of frequency noise of superconducting cavities at these temperatures.

Qualitatively, Fig.~\ref{fig:waterfall}(b--d) demonstrate all the hallmarks of motional narrowing due to one or more RTS fluctuators~\cite{Abragam1961,Borbat2001Jan,Perlow1968Aug,Berthelot2006Oct, Cassabois2010}. At low power, we see multiple frequency positions, %histograms 
which can be attributed to %a finite time sampling of the distribution due to
several slowly varying RTS signals. If we were to continue measuring for even longer time periods, we would ultimately expect a Gaussian distribution of frequency shifts~\cite{Cassabois2010}. As the power is increased, these peaks coalesce into a single distribution whose width narrows as the power increases. To obtain an estimate for the power dependence of this narrowing, we fit the {\small{FWHM}}, shown via the dashed lines in Fig.~\ref{fig:waterfall}(e--g), to the functional form $F_0 + F_1/\left<n\right>^\beta$, and we find a $\beta$ value of 0.58, 0.82, and 0.63 for the nanowire, resonator, and cavity, respectively (see Table~\ref{tab:fig1_fits} and discussion in supplement). 

%%TC:ignore
\begin{table}
    \centering
    \caption{\label{tab:devices} Characteristics of the three resonators}
    
    \begin{ruledtabular}
    \begin{tabular}{cccccc}
        Resonator & $f_r$\,(GHz) & $Z_c\,(\Omega)$ & $Q_i$ & $Q_c$ \\ \colrule
        Nanowire~\cite{Niepce2019Apr}  & 5.3 & $\SI{6.8e3}{}$ & $\SI{2.5e4}{}$ & $\SI{8.0e4}{}$ \\
        Coplanar~\cite{Burnett2018Mar} & 4.3 & 50 & $\SI{5.4e5}{}$ & $\SI{1.8e5}{}$\\
        Cavity~\cite{Kudra2020Jun}  & 6.0 & 58 & $\SI{1.1e7}{}$ & $\SI{8.2e6}{}$ \\
    \end{tabular}
    \end{ruledtabular}
\end{table}
%%TC:endignore

\subsection{Spectral and Allan analysis of fluctuations: universal dependence of individual RTS fluctuators on the applied drive power}

To gain further insight into the fluctuations, we %perform a statistical analysis and 
examine the spectral properties (Fig.~\ref{fig:fits}(a)) and Allan deviation (Fig.~\ref{fig:fits}(b)) of the frequency fluctuations. While the frequency spurs in the time-series data in Fig.~\ref{fig:waterfall}(a) are indicative of RTS noise, the spectral and Allan responses allow us to quantitatively fit the data and identify the unique characteristics of an RTS response~\cite{VanVliet1982Jul}, in contrast to other types of noise (e.g.\@ `white' or `$1/f$'). The data in Fig.~\ref{fig:fits} prominently features a single dominating RTS fluctuator (see Methods, eqs.~(\ref{eq:lor_psd}--\ref{eq:lor_allan}) for the functional form), which we can fit to extract its amplitude $A$, corresponding to a frequency step size between the states of the telegraph noise process, and characteristic time $\tau_0$.

\begin{figure*}[t]
    \includegraphics[width=6.8cm]{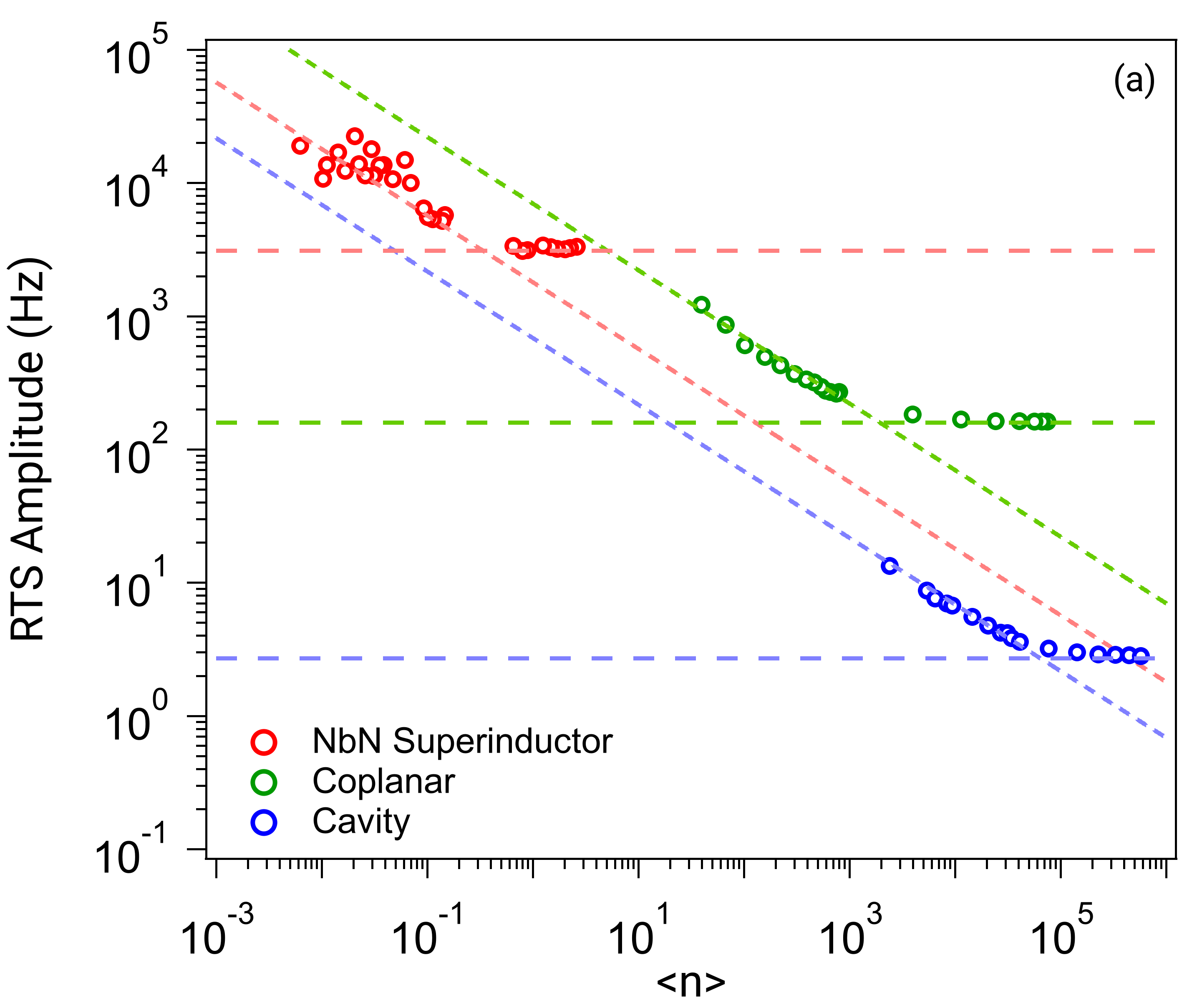} % width changed to 0.8*8.5cm
    \includegraphics[width=6.8cm]{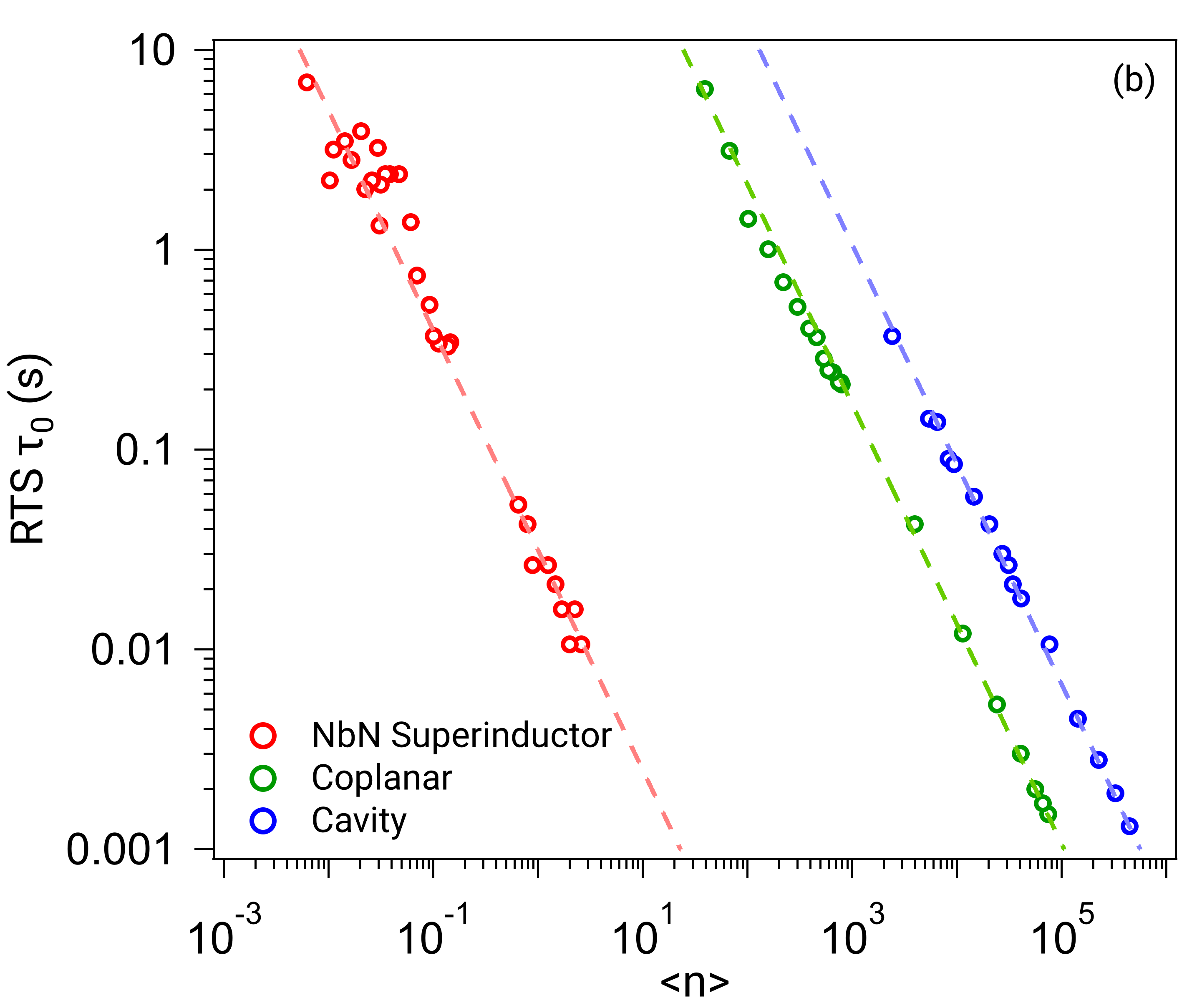} % width changed to 0.8*8.5cm
    
    \caption{\label{fig:summary} Drive-power dependence of the RTS amplitude $A$ \textbf{(a)} and switching time constant $\tau_0$ \textbf{(b)} determined from noise data from the three resonators (Fig.~\ref{fig:sweep}) fitted to the RTS model (Eq.~\ref{eq:lor_allan}). 
    The horizontal dashed lines in (a) indicate the saturation $A\to A_0$, related to the minimum \small{FWHM}\normalsize{} in Fig.~\ref{fig:waterfall}(e--g);
    the diagonal lines in (a) indicate $1/\sqrt{\langle n\rangle}$ scaling (not a fit).
    The dashed lines in (b) are fits of $\tau_0$ to the power law $(\left<n\right>/n_c)^{-\alpha}$ (Eq.~\ref{eq:power_laws_tau}), with $\alpha=1.1$. 
    The fitted parameters are presented in Table~\ref{tab:fits}.
    }
\end{figure*}

We analyse the fluctuation data for a range of drive powers, shown in Supplementary Fig.~\ref{fig:sweep}, and we observe that all three devices present similar noise profiles---featuring one dominant RTS fluctuator---albeit at widely different amplitudes: the nanowire is the noisiest and the cavity is the quietest. 
Generally, there exists other less-prominent RTS features, sometimes at sufficient densities that they sum up to a $1/f$-like trend~\cite{Nugroho2013Apr}. In the limit of few RTS fluctuators, or alternatively in the $1/f$ limit, the data can be reliably fitted. However, between these limits, it becomes non-trivial to determine the exact number of RTS fluctuators that describe the fluctuations. For consistency, we therefore focus on determining the characteristic switching time $\tau_0$ and amplitude $A$ of the dominant RTS fluctuator within our measurement window and plot the resulting values of $A$ and $\tau_0$ vs.\@ $\langle n\rangle$ in Fig.~\ref{fig:summary}(a) and (b), respectively.

%%TC:ignore
%%TC:endignore

When examining the raw frequency jitter (Fig.~\ref{fig:waterfall}(a)), an initial assumption would be that the noise present is a mixture of RTS (on $\sim$100-second timescale) and `white' frequency noise (i.e. $S_{y} \propto f^0$ and $\sigma_{y} \propto \tau^{-0.5}$). However, from the PSD and Allan deviation methods, it is clear that no white frequency noise is present (in the supplemental this is shown for all microwave drives). Therefore, the noise present is a combination of an RTS at timescales of $\sim$100~seconds and other RTS at much smaller timescales $\sim$1~ms to 1~s (see Fig.~\ref{fig:summary}(b)). As such, the multi-peak behaviour of Fig.~\ref{fig:waterfall}(b--d) occurs due to the longer-timescale RTS, whereas the width in Fig.~\ref{fig:waterfall}(e--g) is determined by the smaller-timescale RTS. Within our measurement window, the shorter timescale RTS dominates the signal, from which we extract the parameters $A$ and $\tau_{0}$. 

In Fig.~\ref{fig:summary}(a), we see that $A$ is initially power dependent, decreasing with increasing power. However, it saturates at high powers, starting at a photon number corresponding approximately to the coalescence of peaks in Fig.~\ref{fig:waterfall}(b--d) ($\langle n\rangle\sim 0.1$ for the nanowire and $10^4$ for the cavity; here we emphasise that the conversion from photon occupation to electric field is very different for each resonator). All three devices show this behaviour, although the amplitudes, saturation levels $A_0$ (see Table~\ref{tab:fits}), and the cross-over points vary. 

Furthermore, as shown in Fig.~\ref{fig:summary}(b), we find that the extracted $\tau_0$ values of the three resonators follow an empirical power law
\begin{equation}\label{eq:power_laws_tau}
    \tau_0(\langle n\rangle) = (1\,\mathrm{s})\times(\left<n\right>/n_c)^{-\alpha}
\end{equation}
where $\alpha$ is found close to 1.1 in all three resonators, and $n_c$ is a ``critical" photon number, unique for each device; see the fit parameters in Table~\ref{tab:fits}.

%%TC:ignore
\begin{table}[b]
    \centering
    \caption{\label{tab:fits} Fit parameters for the dominant RTS fluctuators' switching times $\tau_0$ vs.\@ drive power $\langle n\rangle$ (Eq.~\ref{eq:power_laws_tau}) and saturation values ($A_0$) of their amplitudes $A$ for large $\langle n\rangle$, shown in Fig.~\ref{fig:summary}. 
    The FWHM values refer to the histograms in Fig.~\ref{fig:waterfall}(e--g) at high power.}
    %$\tau_0 \propto (\left<n\right>/n_c^{(\tau)})^{-\alpha}$ and 
    %$A = A_0 + (\left<n\right>/n_c^{(A)})^{-\beta}$.

    \begin{ruledtabular}
    \begin{tabular}{ccc}
        Device  & RTS $\tau_0$ & RTS $A$ \\ \colrule
        \multirow[t]{3}{*}{Nanowire}    
        & $\alpha = 1.1$ & $A_0 = \SI{2.8e3}{\hertz}$\\
                        & $n_c = \SI{4.3e-2}{}$  
                        & $\mathrm{FWHM}= \SI{1.2e4}{\hertz}$ \\
                        \colrule
        \multirow[t]{3}{*}{Coplanar}   
        & $\alpha = 1.1$ & $A_0 = \SI{1.6e2}{\hertz}$\\
                        & $n_c = \SI{2.0e2}{}$   
                        & $\mathrm{FWHM}=    \SI{2.7e2}{\hertz}$ \\   
                        \colrule
        \multirow[t]{3}{*}{Cavity}      
                        & $\alpha = 1.1$ & $A_0 = \SI{2.5}{\hertz}$ \\
                        & $n_c = \SI{1.1e3}{}$   
                        & $\mathrm{FWHM}=    \SI{9.4}{\hertz}$ 
    \end{tabular}
    \end{ruledtabular}
\end{table}
%%TC:endignore

\section{Discussion}

%%TC:ignore
\begin{figure*}[t]
    \includegraphics[width=17cm]{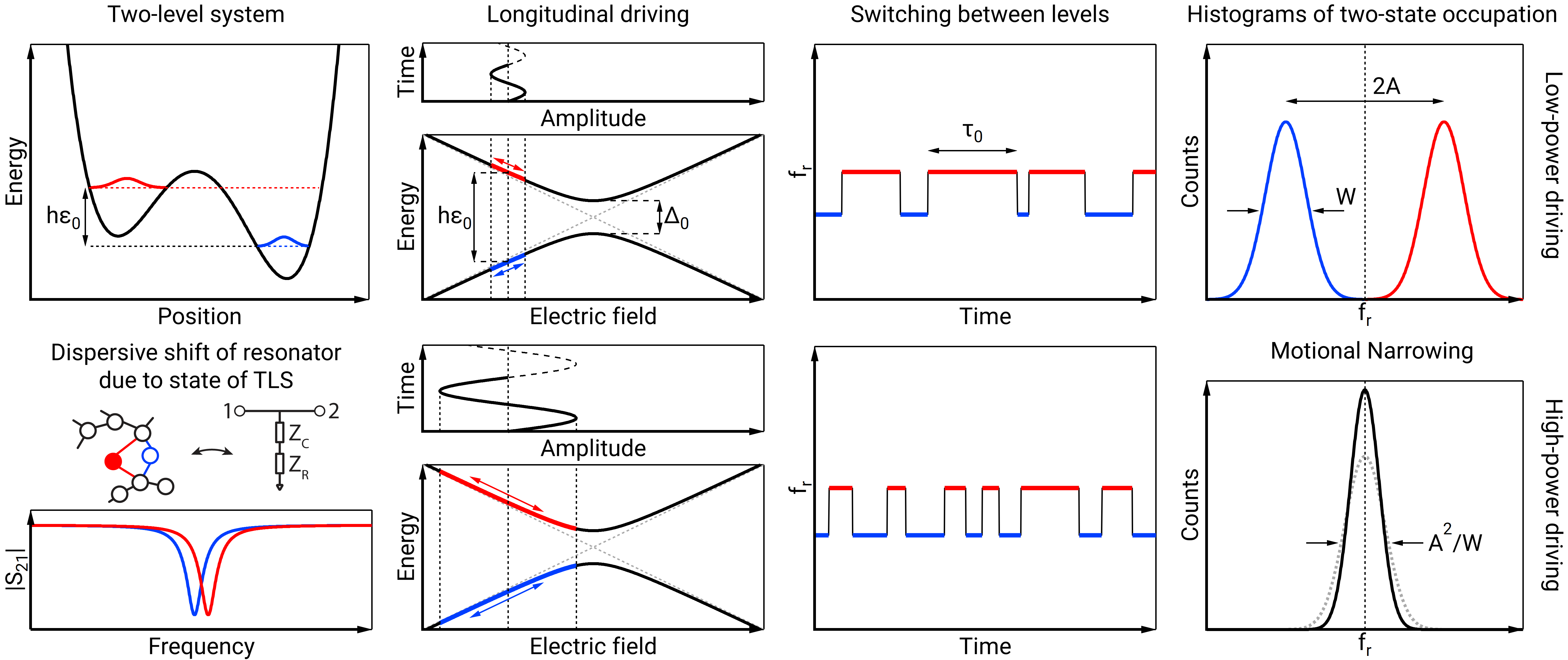} %Note fig size changes to 08*17cm
    \caption{\label{fig:cartoon} An illustration of the relevant RTS switching regimes (high and low power driving) resulting from  small and large-amplitude driving of a TLS about a bias point $\varepsilon_0$ near (but not at) its degeneracy point $\varepsilon=0$. The resulting transitions between the two eigenstates of the TLS result in different dispersive shifts of the resonator, resulting in RTS fluctuations of the resonance frequency.}
\end{figure*}
%%TC:endignore

The power dependence of the histogram width and the noise characteristics revealed by the Allan deviation can be understood in terms of motional narrowing by one or a few dominant RTS fluctuators. We now show how the resonator field can `sympathetically' drive two-level defects in the surrounding dielectric in a regime that results in RTS noise with the required power dependence to explain the observations. This effect of sympathetic driving of the bath of defects and the resulting motional narrowing likely influences the power dependence in many superconducting devices.

\subsection{Motional narrowing}

Together, the plots in Fig.~\ref{fig:waterfall} highlight the power-dependent transition from multi-peaked behaviour at low circulating power in the resonator to single-peaked behaviour at high power. Additionally, as the power increases, the widths of the histograms narrow. Such behaviour is indicative of motional narrowing (motional averaging)~\cite{Abragam1961}, where a multi-level system transitions into a single-level system that also exhibits increased spectral stability. Motional narrowing is a common phenomenon that has been found in a wide variety of systems: NMR spectra~\cite{Abragam1961,Kohmoto1994Jun}, ESR spectra~\cite{Borbat2001Jan}, gamma emissions~\cite{Perlow1968Aug}, superconducting qubits~\cite{Li2013Jan}, and two-level NV-centre defects~\cite{Jiang2008Feb,Bluvstein2019}. However, despite the similarity between an NV centre and a parasitic TLS, motional narrowing has not been considered in the framework of dielectric loss, charge noise, or other TLS phenomena that manifest within superconducting circuits.

The observation of quasi-stable resonant frequencies is consistent with the model of a bath of spectrally unstable, charged TLS that are dispersively coupled to the resonator~\cite{Burnett2014Jun,Faoro2015Jan,Muller2019Oct,Klimov2018Aug}.
In previous studies of resonators, the coupling to many TLS manifested as a $1/f$ noise spectrum~\cite{Burnett2014Jun,Neill2013Aug,deGraaf2018Mar}. Within studies on superconducting qubits, the coupling to TLS has been strong enough to result in an RTS noise  spectrum~\cite{Burnett2019Jun,Schlor2019Nov}. 
The RTS noise behaviour found here demonstrates a similarly strong coupling to single or few individual TLS. 

Typically, in such a model of dispersively coupled (near-resonant) TLS, their dynamics are dominated by incoherent, low-frequency two-state fluctuators whose fluctuations dephase the TLS (widen its spectrum) or shift the TLS energy~\cite{Mueller2009,Burnett2014Jun,Faoro2015Jan,Muller2015Jul}. This results in a $1/f$ noise spectrum which scales as approximately $1/\sqrt{\langle n \rangle}$. 
Here we see single- or few-RTS dynamics rather than $1/f$ noise, where the RTS amplitude scales as $1/\sqrt{\langle n \rangle}$ (Fig.~\ref{fig:summary}(a)) up until some critical power, beyond which is becomes power independent. 
However, a very clear nearly $1/\langle n \rangle$ dependence of $\tau_0$ (Fig.~\ref{fig:summary}(b)) over all powers suggests that the switching rate requires a different interpretation.

To understand the ramifications of the observed power dependence, we consider an RTS system with only two states, at frequencies $\pm A$, with a characteristic switching rate between these states of $W$ per unit time. For slow switching, $W\ll |A|$, the spectral response of the RTS signal consists of two peaks at frequencies $\pm A$ with a width ({\small{FWHM}}) given by $W$. In the opposite limit of strong driving, $W \gg |A|$, the resonance is a single peak centred at zero frequency with {\small{FWHM}} width $A^2/W$, which is narrower. Importantly, motional narrowing can extend beyond the simple two-state to one-state example that we have described~\cite{Abragam1961}; in multiple-state examples~\cite{Borbat2001Jan,Perlow1968Aug}, multiple $W$ and $\pm A$ exist, although the convergence towards a single narrow state still occurs in the strong driving limit~\cite{Berthelot2006Oct, Cassabois2010}, which is the regime we focus on. The observation that $\tau_0\propto 1/\langle n\rangle^{1.1}$ in the fast fluctuation limit therefore suggests that $W \propto \langle n\rangle^{1.1}$, and this observation is common across all three devices.

\subsection{Landau--Zener transitions in the bath of TLS defects}

In order to investigate the mechanism for modulation of the TLS defect by the resonator, and to explain the results presented above, we start from the assumption that the bath of fluctuators driving the RTS behaviour is described by the standard tunnelling model~\cite{Phillips1987Dec,Muller2019Oct}, where each defect can be described by the TLS Hamiltonian
\begin{equation}\label{eq:H_tls}
    \hat{\mathcal{H}}_0= (-h/2) (\varepsilon\hat{\sigma}_z + \Delta_0\hat{\sigma}_x)
\end{equation}
as illustrated in Fig.~\ref{fig:cartoon}.
The tunnel coupling $\Delta_0$ and bias $\varepsilon$ vary from defect to defect and are a function of the local atomic environment. We assume that the electric field of the resonator couples to the defects via their charge dipole, i.e.\@ longitudinally (along $\hat{\sigma}_z$) in the basis of uncoupled double wells. The bias is therefore comprised of a constant offset and a time dependent term, 
\begin{equation}
    \varepsilon(t) = \varepsilon_0 + \varepsilon_\mathrm{rf} \cos(2\pi f_r t),
\end{equation}
where $\varepsilon_\mathrm{rf}$ has units of frequency but is proportional to the amplitude of the resonator electric field ($\propto\!\!\sqrt{\langle n\rangle}$), and hence to the radio-frequecy (rf) voltage at the source.
%\sout{and frequency sweep rate $v=\mathrm{d}\varepsilon/\mathrm{d}t$} 

For low-loss devices, there are relatively few defects with values of $\Delta_0$ close to the resonator frequency~\cite{Lisenfeld2015Feb,Klimov2018Aug}; however, that is not the parameter regime we are considering. There are also TLS whose $\Delta_0$ is relatively small, but whose equilibrium position (given by $\varepsilon_0$) is such that their eigenstates are nearly resonant with the resonator (see Fig.~\ref{fig:cartoon}). For large resonator fields, the drive can sweep the fluctuator through the TLS anticrossing ($\varepsilon_\mathrm{rf}\approx \varepsilon_0$) or at least near it. We therefore need to consider the role of Landau-Zener (L-Z) tunnelling which can result in transitions between the ground and excited states of the TLS~\cite{Oliver2009, Oliver2005Dec}. 

We can rewrite the Hamiltonian (\ref{eq:H_tls}) above in a time-dependent rotating frame to obtain
\begin{equation}\label{eq:H_rf}
    \hat{\mathcal{H}}_{RF}= (-h/2) (\delta\hat{\sigma}_{\tilde{z}} + \Delta_0 J_1(\lambda)\hat{\sigma}_{\tilde{x}}),
\end{equation}
where $\delta=\varepsilon_0-f_r$ is the detuning between drive and frequency splitting at the bias point $\varepsilon = \varepsilon_0$, $J_1(\lambda)$ is the first-order (one photon) Bessel function of the first kind, representing a dressed gap, and $\lambda = \varepsilon_\mathrm{rf}/f_r$ is the ratio of driving amplitude to driving frequency~\cite{Oliver2009}.

The relevant regime of L-Z driving of TLS in the dielectric of the resonators is that the effective transition rate $W$ between states is less than the dephasing rate ($\Gamma_2$) but greater than the relaxation rate ($\Gamma_1$), i.e.\@ $\Gamma_1 < W < \Gamma_2$. 
In this regime, at resonance ($\delta=0$) in the small-amplitude drive limit ($\varepsilon_\mathrm{rf}\ll f_r$), the one-photon transition rate between the eigenstates is given by~\cite{Berns2006}
\begin{equation}\label{eq:LZ_quasiclassical_rate}
    W(\lambda) = \frac{\pi^2}{2} \frac{\Delta_0^2 \lambda^2}{\Gamma_2}.
\end{equation}
Now, as there is little to no coherence between the two eigenstates, we can consider $W(\lambda)$ as the RTS switching rate, i.e., $\tau_0 = 1/W(\lambda)$, which means that $\tau_0\propto 1/\varepsilon_\mathrm{rf}^2 \propto 1/\langle n\rangle$, where the proportionality constant ($n_c$ in Eq.~\ref{eq:power_laws_tau}) is a product of three unknowns: the decoherence rate, the energy splitting, and the electric-field amplitude at the site of the TLS.

We note that our observed transition rate has a small additional contribution as the amplitude is increased (cf.\@ the exponent $\alpha=1.1$ in Eq.~(\ref{eq:power_laws_tau}) found empirically for all three resonators). We may attribute this to the TLS having a sufficiently large response to the resonator field that higher photon number transitions are non-negligible.

The role of Landau--Zener driving of TLS in the dielectric of qubits and resonators has been previously studied~\cite{Matityahu2019, Burin2013Apr, Khalil2014Sep}; however, in such experiments the mechanism is modulating the frequency splitting of near-resonant TLS as they traverse the resonator frequency, thereby driving non-adiabatic Landau--Zener transitions. The transitions we consider (away from the degeneracy point) influence the dephasing noise (i.e.\@ the low-frequency, real part of the spectral function), similarly to Ref.~\cite{Bluvstein2019}, whereas Ref.~\cite{Matityahu2019} deals with the loss (i.e.\@ the near-resonant, imaginary part leading to energy relaxation). 

\subsection{The role of the ensemble}

While this picture explains the common response between devices and the power dependence of $\tau_0$, it does not explain the low-power response of $A_0$ nor the `more conventional' (but less universal) response of the {\small{FWHM}}. However, both can be explained in terms of the ensemble of RTS fluctuations stemming from multiple TLS. As the power is reduced, below the point of coalescence in the motional narrowing picture, the fit to a single RTS fluctuator no longer captures the key characteristics of the response. The contributions from both additional RTS sources and other noise processes start to dominate and this results in an additional power dependence to the noise amplitude. The diagonal lines in Fig.~\ref{fig:summary}(a) represent a $1/\sqrt{\langle n \rangle}$ scaling, which one would typically expect for $1/f$ noise, indicating that at lower powers, the ensemble response is more dominant. Similarly, the extracted {\small{FWHM}} in Fig.~\ref{fig:waterfall}(e--g) is a function of the entire spectrum, which includes both additional (non-TLS) processes and contributions due to the TLS-TLS interactions in the bath~\cite{Faoro2012Oct,Faoro2015Jan,Kirsh2017Jun,Burin2018Jun}. As these contributions depend on the density and interaction strength between the TLS, they vary more between devices resulting in the differing power response; cf.\@ Table~\ref{tab:fig1_fits} in Supplement. 

\section{Conclusion}

We have studied the frequency noise of three commonly used superconducting resonators within the same measurement and analysis infrastructure. %experimental setup.
We find that in all devices, the noise is described by an RTS process, %a Lorentzian process, 
which we attribute to %RTS caused by 
spectrally unstable TLS.
When studying the %noise behaviour of single 
RTS behaviour with microwave drive power, we find that the switching times follow a common scaling across all types of resonators. 
We interpret the power dependence of the RTS switching times in terms of sympathetic driving of TLS defects by the resonator field. This driving induces Landau-Zener-type resonant transitions, even for TLS whose equilibrium configuration is relatively detuned from the degeneracy point between the two states.
Fundamentally, this highlights the power of standardised testing with common methods. Here, the ability to directly compare different types of superconducting resonator has revealed a commonality of the dominant noise process.
%This fundamentally shows that the dominant noise processes are common across these different superconducting materials and geometries. 
These findings expand the toolkit and material parameter range for examining parameter fluctuations, which has become the leading problem in superconducting quantum-computing efforts. Furthermore, the studies of the nanowire superinductor are particularly relevant to the rapidly growing area of high-impedance qubits~\cite{Niepce2019Apr,Grunhaupt2019Apr,nguyen2019high,hazard2019nanowire}. 

%%TC:ignore
\section{Methods}

\subsection{Device characteristics} 

The examined resonators have similar resonant frequencies, but otherwise have very different superconducting properties and microwave electric-field distributions.
The superinductor consists of a disordered NbN nanowire with high kinetic inductance, and consequently high characteristic impedance $Z_c$, on a Si substrate. %, with a resonant frequency for the measured full-wave mode of $f_r = \SI{5.3}{\giga\hertz}$, characteristic impedance $Z_c = 6.8~\mathrm{k}\Omega$, and $Q_i = \SI{2.5e4}{}$ at single-photon excitation, $\langle n\rangle=1$. It is thoroughly described in Ref.~\cite{Niepce2019Apr}.
The coplanar waveguide resonator was made of Al on Si. %Its fundamental $f_r = \SI{4.3}{\giga\hertz}$, $Z_c = 50~\Omega$, and $Q_i=\SI{5.4e5}{}$ at $\langle n\rangle=1$. Details are found in Ref.~\cite{Burnett2018Mar}. 
The stub-geometry 3D cavity was machined out of 4N-grade Al. %and has $f_r = \SI{6.0}{\giga\hertz}$, $Z_c = 58~\Omega$, and $Q_i=\SI{1.1e7}{}$ at $\langle n\rangle=132$ (the lowest number measured). Details are found in Ref.~[Kudra2020].
The device characteristics of the three resonators are summarised in Table~\ref{tab:devices}, and their designs and fabrication techniques are thoroughly described in the given references. 

The internal quality factors, $Q_i$, of the nanowire and the coplanar waveguide were determined at an average photon occupation number of $\langle n\rangle=1$, whereas that of the cavity was determined at $\langle n\rangle=132$ (the lowest measured); in all cases, this photon occupation corresponds to when $Q_{i}$ has saturated to a low level, consistent with the depolarization of two-level defects.
We determine $\langle n\rangle$, knowing the applied drive power $P$ and the $Q_i$ at that power, $Q_i(P)$, using the relation 
\begin{equation}\label{eq:resonator_fq}
    h f_r \langle n\rangle =  Z_0 Q_l^2 P/\pi^2 Z_c Q_c f_r.
\end{equation}
Here, $h$ is Planck's constant, $f_r$ is the resonant frequency, $Z_0 = 50\,\Omega$ is the impedance of the feedline, and $Q_c$ and $Q_l$ are the coupling and loaded quality factors, respectively, with 
$Q_l^{-1}=Q_c^{-1}+Q_i^{-1}(P)$.

\subsection{Measurement techniques}

The nanowire and coplanar resonators each exhibit a resonance dip due to coupling to a microwave transmission line. The use of a circulator at the cavity input leads to the cavity also exhibiting a resonance dip. The Pound frequency-locked loop (P-FLL) is locked to this resonance dip. We measure the resonant-frequency fluctuations by sampling the frequency of the P-FLL voltage-controlled oscillator using a frequency counter (Keysight 53230A) at a sampling rate of either $\SI{100}{\hertz}$ or $\SI{4}{\kilo\hertz}$. Each noise trace consists of $\SI{1e6}{}$ samples. In addition, once per noise trace, the absolute frequency and microwave power of the signal going into the cryostat are measured with a spectrum analyzer (Agilent E4440A). During a measurement, the cryostat temperature is held constant and noise traces are recorded at various inbound microwave powers.
A detailed description of these measurement techniques is found in Refs.~\cite{Niepce2019Apr,Lindstrom2011Oct}. 

\subsection{Spectral and Allan analysis of fluctuations}

The same raw frequency fluctuations data is used to produce the spectrum of frequency fluctuations $S_y(f)$, using the Welch power spectral density (PSD) estimate with 50\% overlap and a Hamming window, and the overlapping Allan deviation $\sigma_y(\tau)$. A detailed description of this data analysis technique is given in Refs.~\cite{Burnett2019Jun,Rubiola2008}.

The spectral response of a single RTS fluctuator is given by
\begin{equation}
    \label{eq:lor_psd}
    S_y(f) = \dfrac{4 A^2 \tau_0}{1+(2\pi f \tau_0)^2}
\end{equation}
where $A$ and $\tau_0$ denote the RTS amplitude and characteristic time, respectively. The corresponding Allan deviation is given by Ref.~\cite{VanVliet1982Jul}:
\begin{equation}
    \label{eq:lor_allan}
    \sigma_y(\tau) = \dfrac{A\tau_0}{\tau} \left( 4 e^{-\tau/\tau_0} - e^{-2\tau/\tau_0} + 2\dfrac{\tau}{\tau_0} - 3 \right)^{1/2}
\end{equation}
A key strength of the Allan analysis is that it often allows the identification of $\tau_0$ against the noise background, although we use the same parameters when fitting $S_y(f)$ and $\sigma_y(\tau)$.

\subsection{Estimate of errors}

In the determination of $\tau_0$ and $A$ (circles in Fig.~\ref{fig:summary}), we estimate the two standard deviations error to be about 4\% (10\%) for $\tau_0$ (for $A$) for the coplanar and cavity resonators, and for the nanowire resonator at high powers. For the low-power data of the nanowire resonator, we estimate about a factor of two error in both $\tau_0$ and $A$. The collection of longer sets of data would reduce the error.

\section*{Acknowledgements}

We gratefully acknowledge useful discussions with A.\@ Danilov, P.\@ Delsing, and S.\@ Kubatkin. This research has been supported by funding
from the Swedish Research Council, Chalmers Area
of Advance Nanotechnology, and the Wallenberg Center for Quantum Technology (WACQT). JHC is supported by the Australian Research
Council Centre of Excellence program through Grant
number CE170100026 and the Australian National Computational Infrastructure facility. JJB acknowledges financial support from the Industrial Strategy Challenge Fund Metrology Fellowship as part of the UK governments Department for Business, Energy and Industrial Strategy. 

%merlin.mbs apsrev4-1.bst 2010-07-25 4.21a (PWD, AO, DPC) hacked
%Control: key (0)
%Control: author (72) initials jnrlst
%Control: editor formatted (1) identically to author
%Control: production of article title (-1) disabled
%Control: page (0) single
%Control: year (1) truncated
%Control: production of eprint (0) enabled
%

% ****** Supplementary starts here ******

\onecolumngrid
\clearpage
\begin{center}
\textbf{\large Supplemental Material}
\end{center}

%%%%%%%%%% Prefix a "S" to all equations, figures, tables and reset the counter %%%%%%%%%%
\setcounter{section}{0}
\setcounter{equation}{0}
\setcounter{figure}{0}
\setcounter{table}{0}
\setcounter{page}{1}
\makeatletter
\renewcommand{\thesection}{S\Roman{section}}
\renewcommand{\thetable}{S\arabic{table}}
\renewcommand{\theequation}{S\arabic{equation}}
\renewcommand{\thefigure}{S\arabic{figure}}
\renewcommand{\bibnumfmt}[1]{[S#1]}
\renewcommand{\citenumfont}[1]{S#1}
%%%%%%%%%% Prefix a "S" to all equations, figures, tables and reset the counter %%%%%%%%%%

\section{Spectral and Allan analysis of the fluctuations data vs.\@~power for all three resonators}

\begin{figure*}[h]
    \includegraphics[width=5cm]{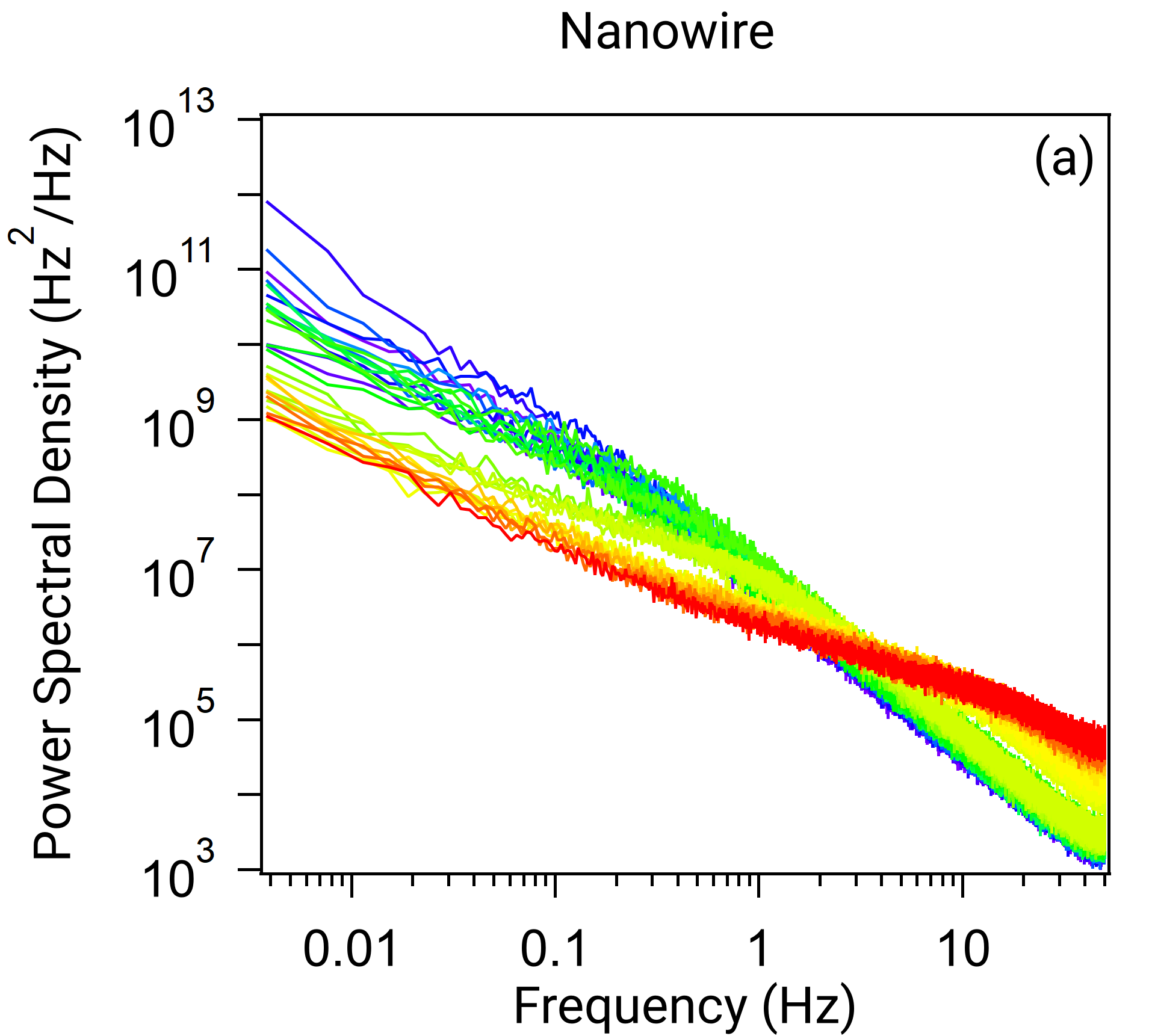} % width changed to 0.8*5.6cm
    \includegraphics[width=5cm]{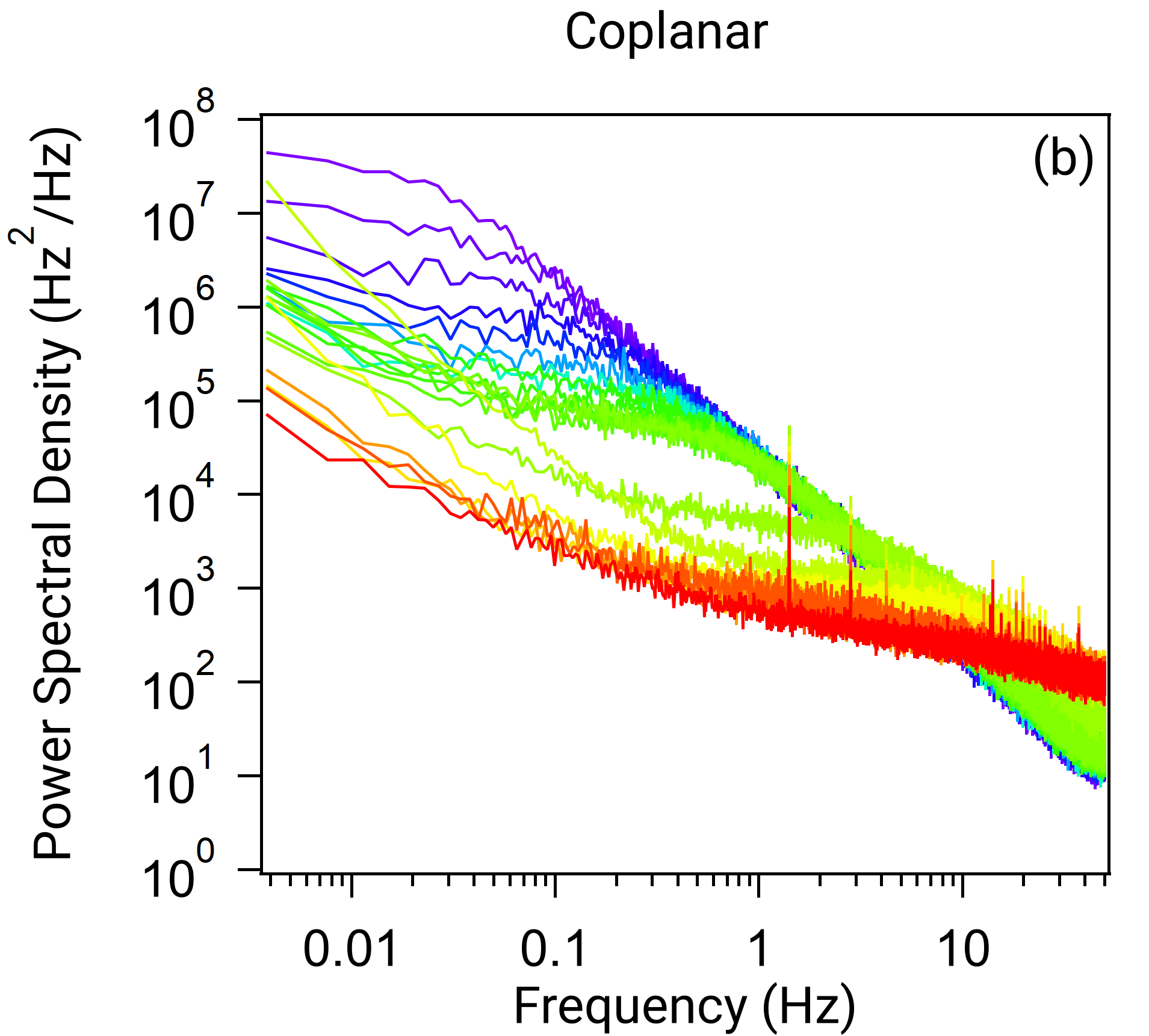} % width changed to 0.8*5.6cm
    \includegraphics[width=5cm]{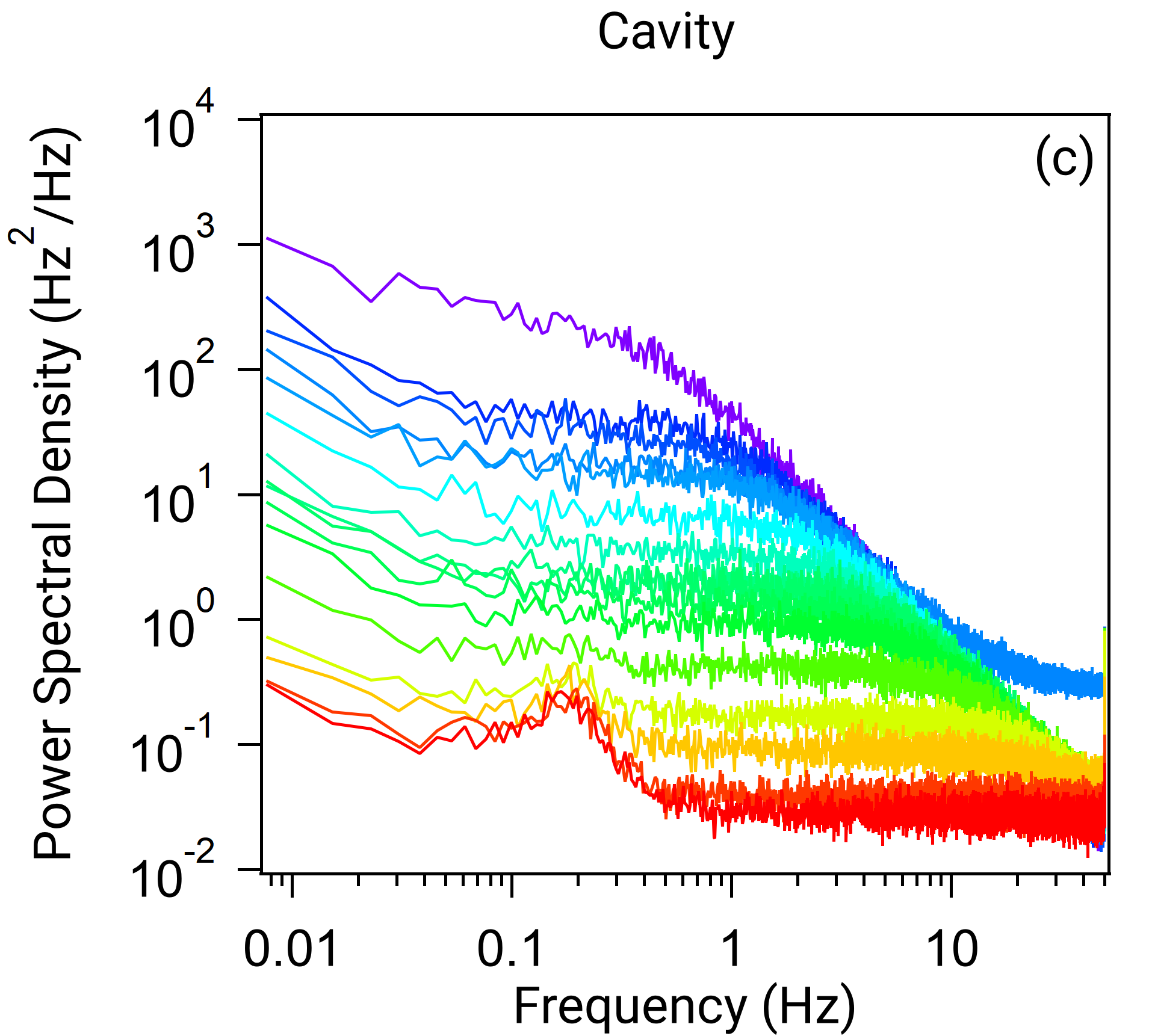} % width changed to 0.8*5.6cm
    
    \includegraphics[width=5cm]{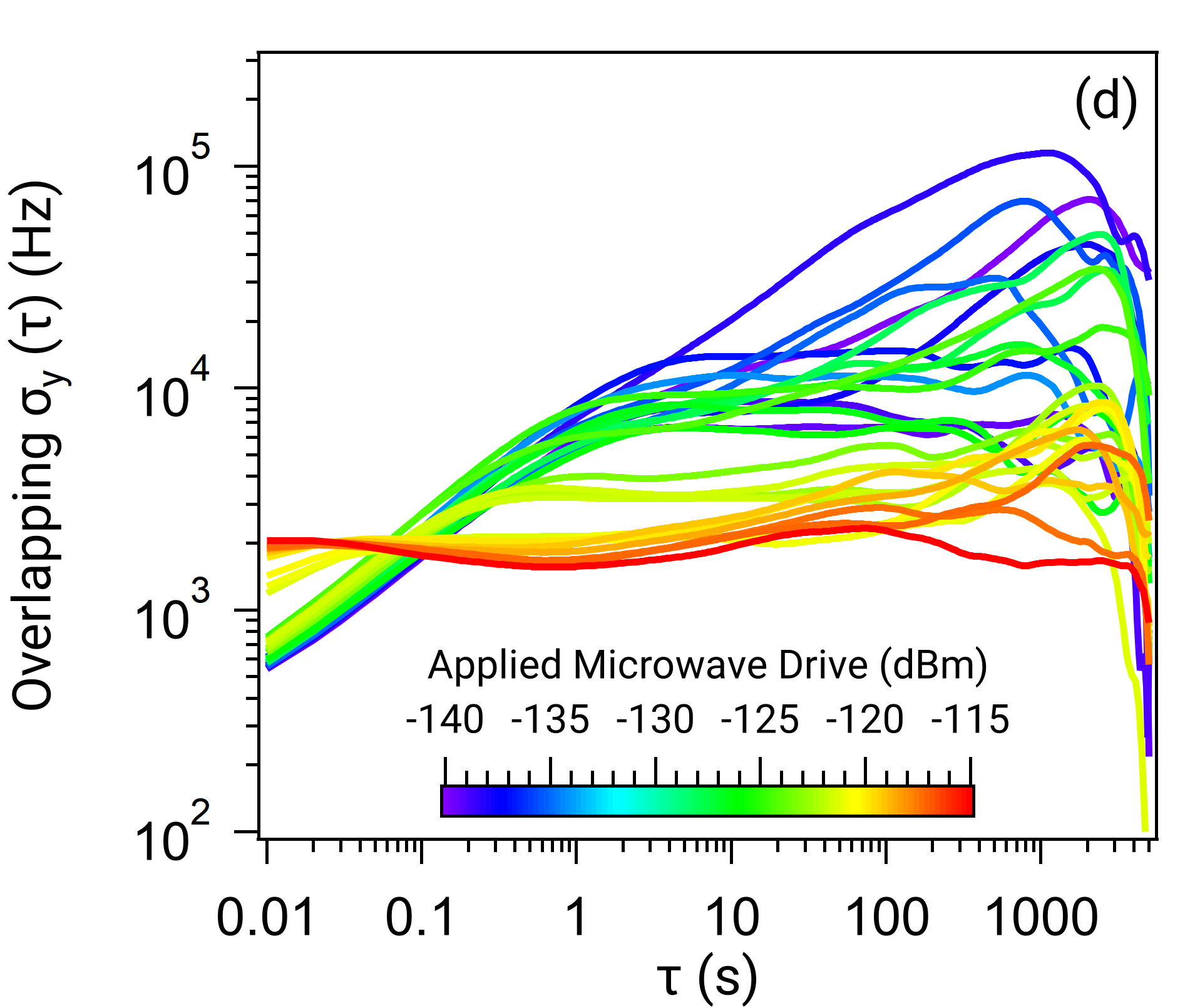} % width changed to 0.8*5.6cm
    \includegraphics[width=5cm]{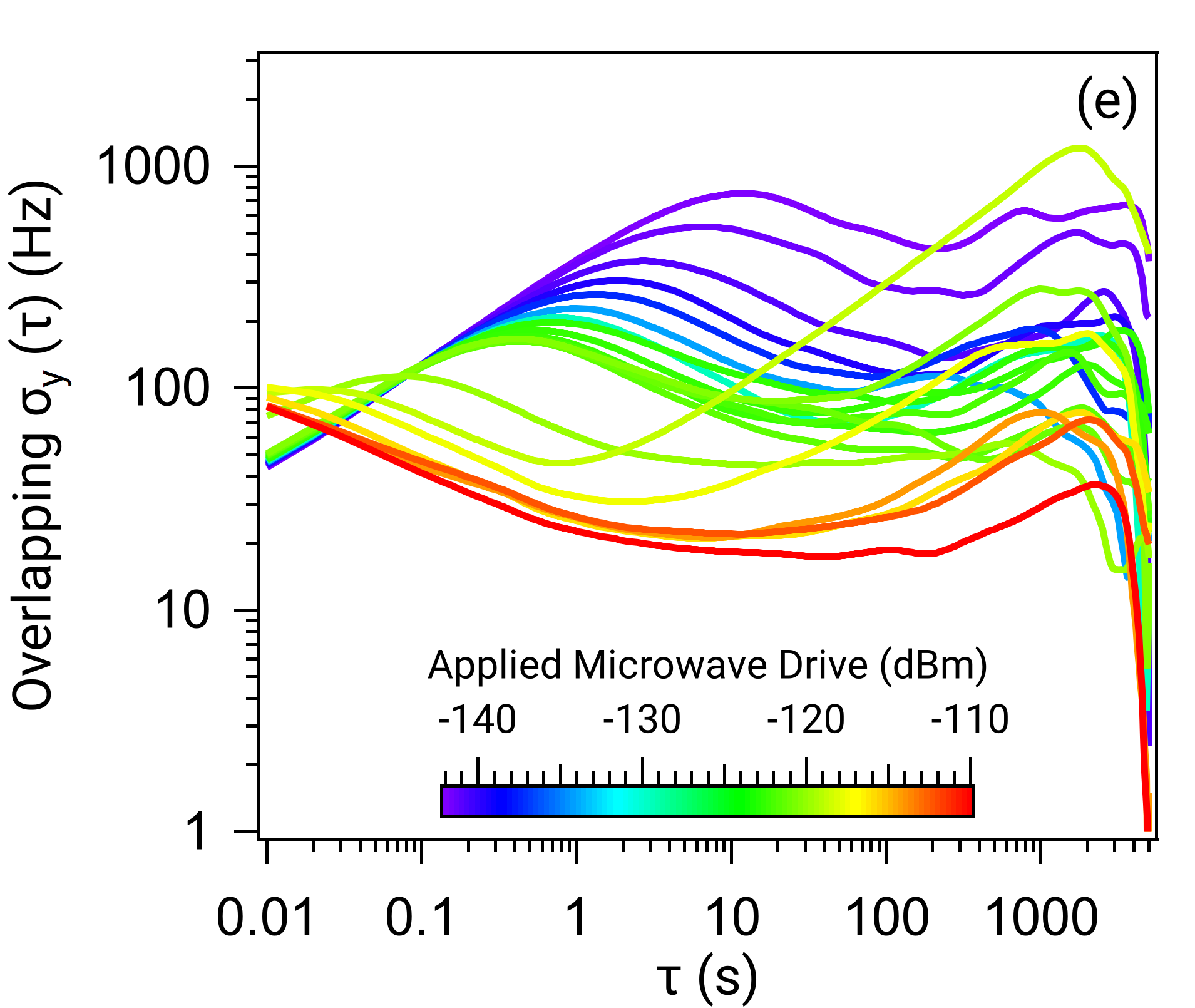} % width changed to 0.8*5.6cm
    \includegraphics[width=5cm]{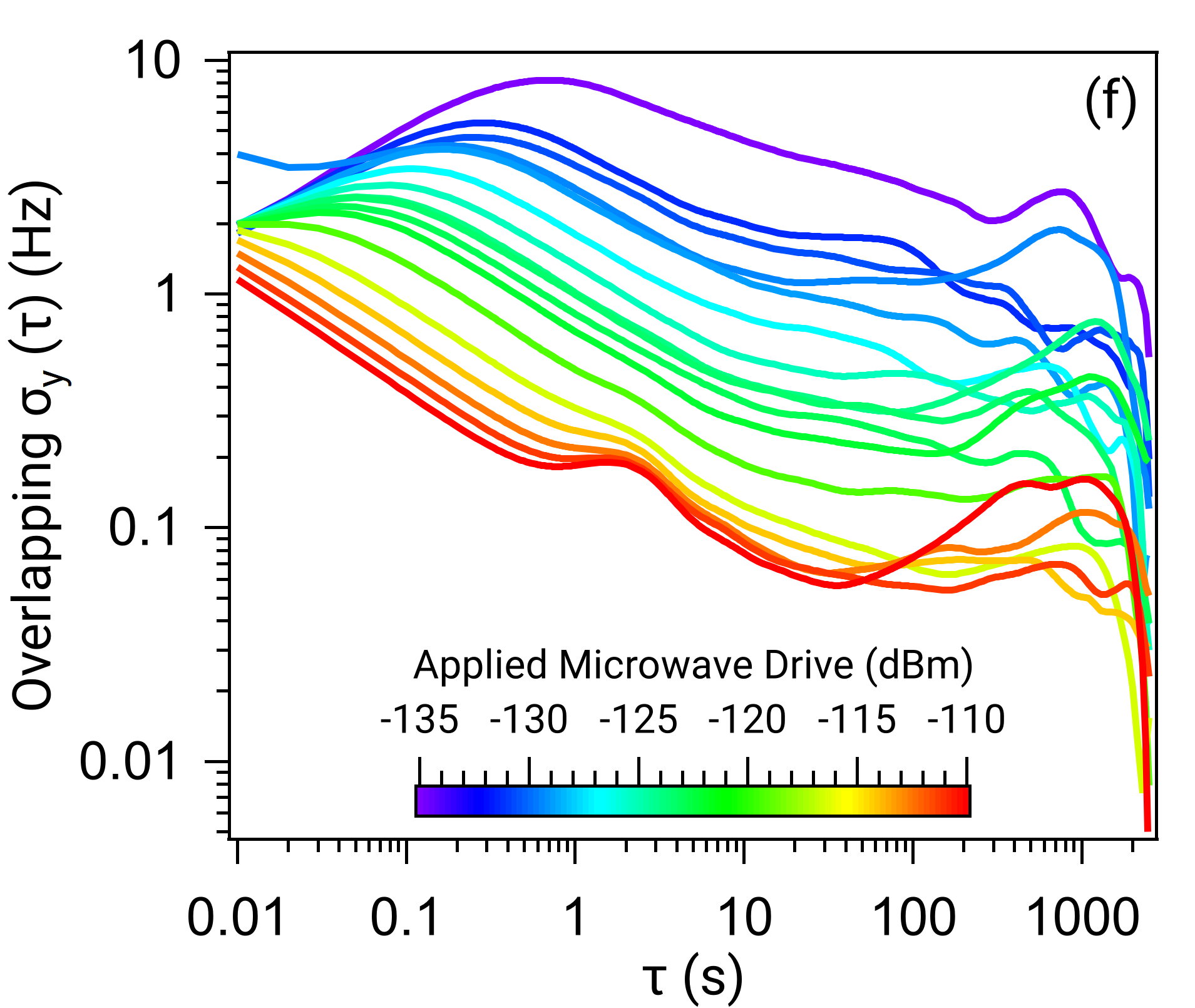} % width changed to 0.8*5.6cm
    
    \caption{\label{fig:sweep} Noise analysis vs.\@ applied power for the three resonators.
    Plots \textbf{(a--c)} show the Welch power spectral densities, and plots \textbf{(d--f)} show the  overlapping Allan deviations of the frequency fluctuations, for various applied powers. All the data was measured at $T = \SI{10}{\milli\kelvin}$ and a sampling rate of $\SI{100}{\hertz}$. 
    We observe that all three devices present similar noise profiles---featuring one dominant Lorentzian---albeit at widely different amplitudes: the nanowire is the noisiest and the cavity is the quietest. 
    As the microwave drive is increased, we observe that the dominant Lorentzian feature in the PSD (Allan) is consistently moving toward higher frequencies (lower $\tau$) for all devices.
    The cavity resonator frequency is so stable that at high power, its noise falls below that of the reference of our frequency-locked loop, which shows up as an additional peak in the PSD at around 0.2~Hz.}
\end{figure*}

For completeness, here we include the calculated PSD and Allan deviations for all measured powers of each device. In Fig.~\ref{fig:sweep}, the top row shows the calculated PSD, and the bottom row shows the calculated overlapping Allan deviation.  The functional form of the RTS feature was shown in Fig.~\ref{fig:fits} in the main text. Here, the RTS switching time is clearly shown to move towards higher times as the microwave amplitude is decreased.

\section{Power dependence of the resonator response width}
A key method for understanding the influence of decoherence mechanisms in superconducting devices is to map the power dependence of their response.
The standard tunnelling model predicts that the Q factor of a resonator scales as $\sqrt{1+\langle n \rangle/n_c}$, for some critical photon number $n_c$; however, variations from this scaling are often seen in experiments. Therefore, several authors have fitted the power dependence of Q to $\sqrt{1+(\langle n \rangle/n_c)^\beta}$, where $\beta=1$ corresponds to the STM prediction, but typically $\beta<1$ is observed~\cite{sMacha2010Feb,sWisbey2010Nov,sPaik2010Feb,sBurnett2014Jun,sBurnett2017Jul,sRomanenko2017Dec,Kudra2020Jun}. Deviations from the STM-predicted scaling have been interpreted as evidence for TLS-TLS interactions~\cite{sFaoro2012Oct,sFaoro2015Jan,sKirsh2017Jun,sBurnett2014Jun,sBurin2018Jun}, in which case such variations between devices would be considered unsurprising. However, in this experiment, it is the frequency jitter %relative to a phase-locked reference 
that is measured, not the Q factor, and so it is unclear if one would expect a similar response, although the $1/f$ noise has been found to scale with the loss tangent (i.e. with $1/\sqrt{1+\langle n \rangle/n_c}$)~\cite{sBurnett2014Jun,sFaoro2015Jan}. To compare to previous work on Q factors, we use a similar power-law expression to fit to the FWHM of the histograms in Fig.~\ref{fig:waterfall}(e-g), namely $F_0 + F_1/\left<n\right>^\beta$. The resulting fit values are given in Table~\ref{tab:fig1_fits}, showing typical values $0.5<\beta<1$, but which vary from device to device. Such variation between devices is considered `normal' in the literature and serves to highlight how surprisingly similar the scaling of the single-RTS switching rate (Fig.~\ref{fig:summary}(b)), revealed by the Allan analysis, is across all three devices measured.

\begin{table}[H]
    \centering
    \caption{\label{tab:fig1_fits} Fitting parameters for Fig.~\ref{fig:waterfall}(e--g).}
    
    %\begin{ruledtabular}
    %\begin{tabular}{cccc}
    \begin{tabular}{p{0.1\textwidth}>{\centering}p{0.1\textwidth}>{\centering}p{0.1\textwidth}>{\centering\arraybackslash}p{0.1\textwidth}}
            \hline\hline
            Device   & $F_0 \, (\SI{}{\hertz})$ & $F_1 \, (\SI{}{\hertz})$ & $\beta$ \\
            \hline
            Nanowire & $\SI{1.3e4}{}$        & $\SI{5.5e3}{}$        & $\SI{0.58}{}$ \\
            Coplanar & $\SI{4.2e2}{}$        & $\SI{6.4e4}{}$        & $\SI{0.82}{}$ \\
            Cavity   & $\SI{2.6}{}$          & $\SI{1.2e3}{}$        & $\SI{0.63}{}$ \\
            \hline\hline
    \end{tabular}
    %\end{ruledtabular}
\end{table}

%merlin.mbs apsrev4-1.bst 2010-07-25 4.21a (PWD, AO, DPC) hacked
%Control: key (0)
%Control: author (72) initials jnrlst
%Control: editor formatted (1) identically to author
%Control: production of article title (-1) disabled
%Control: page (0) single
%Control: year (1) truncated
%Control: production of eprint (0) enabled
%

%%TC:endignore

\end{document}